\documentclass[prl, byrevtex, twocolumn, showpacs]{revtex4}

\usepackage{graphicx}
\newcommand{\degree}{$^{\circ}$}

\begin{document}

\author{Yuri Zuev, Mun Seog Kim}
\thanks{Present Address: Division of Electromagnetic Metrology, Korea Research Institute of Standards and Science, P.O. Box 102, Yuseong, Daejeon, 305-600, Republic of Korea, e-mail: msk2003@kriss.re.kr} 
\author{Thomas R. Lemberger}
\affiliation{Department of Physics, Ohio State University\\ 174 W. 18th Ave., Columbus, OH, 43210}

\title{Correlation between superfluid density and $T_C$ of underdoped YBa$_2$Cu$_3$O$_{6+x}$ near the superconductor-insulator transition}

\begin{abstract}
We report measurements of the ab-plane superfluid density $n_s$ (magnetic penetration depth, $\lambda$) of severely underdoped films of YBa$_2$Cu$_3$O$_{6+x}$, with $T_C's$ from 6 to 50 K. $T_C$ is not proportional to $n_s(0)$; instead, we find $T_C \propto n_s^{1/2.3 \pm 0.4}$. At the lowest dopings, $T_C$ is as much as 5 times larger than the upper limit set by the KTB transition temperature of individual $CuO_2$ bilayers.
\end{abstract}


\maketitle

The problem of high-$T_C$ superconductivity at severe underdoping is complicated by admixtures of different physics, intrinsic or extrinsic to superconductivity itself, e.g., stripes, pseudogap, metal-insulator transition. In particular, the relationship between the pseudogap and superconductivity is perhaps the central issue in the field~\cite{PLee}. There is now a wide variety of theories available that attempt to explain the coexistence of the pseudogap and superconductivity. At the mean-field level, most of them fail to account for the observed decrease of $T_C$ to zero with underdoping, and they appeal to thermal phase fluctuations to do the job.~\cite{EmeryKivelsonPRL, EmeryKivelsonNature, Carlson1, CurtyBeck, IoffeMillis, Muller, Scalapino, Timm, Hirschfeld, Tesanovic}. This is reasonable if interlayer coupling in underdoped cuprates is weak enough that samples are quasi-2D, and $T_C$ is approximately the 2D-XY (or, KTB~\cite{KTB}) transition temperature. After all, this relationship holds, at least approximately, for cuprates that are moderately underdoped~\cite{Uemura}. 

In some models, ~\cite{EmeryKivelsonPRL, EmeryKivelsonNature, Carlson1} the pseudogap arises from quasi-2D phase fluctuations. Electron pairing occurs at the pseudogap temperature, which increases with underdoping, but phase fluctuations delay phase coherence to a much lower temperature, the observed transition temperature. In this framework, the measured $T_C$ is approximately the KTB transition temperature for a single superconducting layer. Thus, $T_C \propto n_s(0)$. Therefore, the empirical Uemura proportionality between $T_C$ and $n_s(0)$ finds a natural explanation.

In this paper we show that $T_C$ is not proportional to superfluid density for severe underdoping. Near the superconductor-to-insulator transition, $T_C$ is roughly proportional to $n_s(0)^{1/2}$, implying that phase fluctuations do not suppress $T_C$ as has been conjectured.

Experimental progress in this area is impeded mostly by the absence of severely underdoped specimens of high quality. High quality in this case would mean high degree of homogeneity and consequently narrow superconducting transition. The problem is that because the slope $T_C$ vs. $x$ is large near the superconductor-to-insulator transition, small oxygen composition variations across the sample can cause a wide transition. We have made progress on reducing transition widths to where the conclusions of this paper are insensitive to them.

To optimize oxygen homogeneity, we grew our YBCO films sandwiched between two PrBa$_2$Cu$_3$O$_{6+x}$ (PBCO) layers, so that there is no free surface of the YBCO anywhere and no surface barrier for oxygen diffusion into YBCO. PBCO/YBCO/PBCO trilayers were deposited on (001) SrTiO$_3$ substrates by pulsed laser ablation with a Kr-F excimer laser (Lambdaphysik 305i, 248 nm wavelength, pulse energy 150 mJ). For the first PBCO layer, 10 unit cells thick, the heater was at 820\degree C and oxygen pressure was 140 mTorr. After deposition, this layer was fully oxidized at 500\degree\ C for 10 minutes in 760 torr O$_2$. Then a 20 or 40 unit cell (235 or 470\AA respectively) layer of YBCO and 20 or 40 unit cell cap layer of PBCO were deposited at 760 \degree C and 140 mTorr of O$_2$. 

After deposition, the films were annealed {\em in situ} for 12 to 24 hrs in 10 to 200 torr O$_2$ at 600\degree C or 700\degree C and then either quenched by dropping them onto crumpled aluminum foil or slowly cooled with the heater turned off. It took about an hour to cool from 600\degree C to under 200\degree C, where oxygen kinetics becomes negligibly slow. Our samples had $T_C$'s down to 6 K, and even at this low $T_C$ the peak in $\sigma_1$ was well-defined, if broad, whereas in previous attempts the peak in $\sigma_1$ extended nearly to $T = 0$.

Samples annealed for 12 hours showed lower $T_C$ and superfluid density than those annealed for 24 hours at the same (or even slightly lower) oxygen pressure and same temperature. From this we conclude that 12 hours is not enough to reach equilibrium with atmosphere. For the same anneal time, lower oxygen pressure results in lower $T_C$. From $T_C$ we infer oxygen content $x$ using a canonical phase diagram (see, e.g.~\cite{Jorgensen}). The films have $0.37\leq x \leq 0.53$.

From our mutual inductance measurements at about 50 kHz we determine the sheet conductance $\sigma d_{\mathrm {film}}=\sigma_1d_{\mathrm {film}}-i\sigma_2d_{\mathrm {film}}$ of the film. From the imaginary part, $\sigma_2d_{\mathrm {film}}$, we extract the magnetic penetration depth $\lambda^{-2}=\sigma_2\mu_0\omega$, and hence the superfluid density, $n_s \propto \lambda^{-2}$. The details are given in references~\cite{2coil1,2coil2}. Inhomogeneities in oxygen content contribute to the natural width of the peak at $T_C$ in the real part of the conductivity, $\sigma_1$. It is this spread that was in the way of observing subtle details in the data near $T_C$. Our best samples have full width of this peak of $\approx 2K$, while others have several Kelvin wide peaks.

All measurements for this study were performed in doubly $\mu$-metal shielded
cryostats to eliminate the effect of the Earth's magnetic field. For
comparison, one data set was taken in both shielded and unshielded
cryostats. The latter showed slightly lower $T_C$, therefore the $\mu$-metal shields were an important precaution.

\begin{figure}[t]
%\centering
\includegraphics[width=\columnwidth]{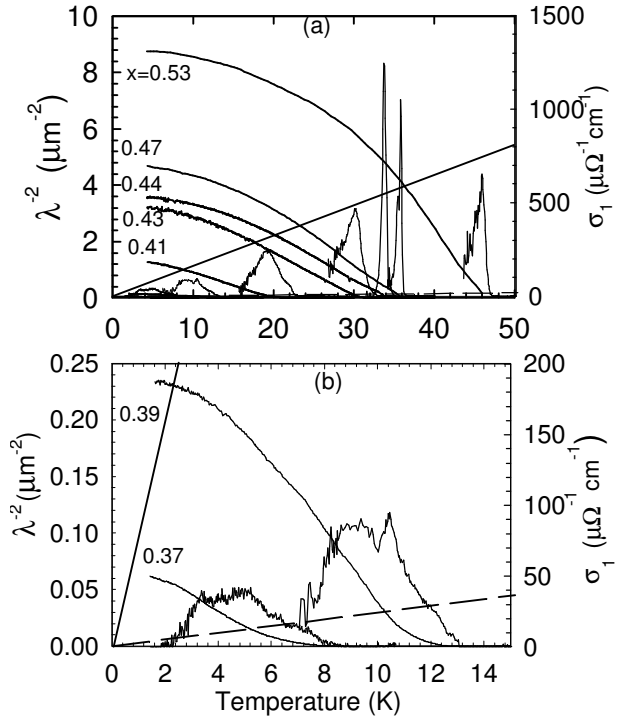}
\caption{(a) Superfluid density $n_s\propto \lambda^{-2}$ and real conductivity $\sigma_1$ as functions of $T$ for YBa$_2$Cu$_3$O$_{6+x}$ films with oxygen contents: $x=$ 0.53, 0.47, 0.43, 0.42, and 0.41; (b) Close-up of 0-15 K region for films with $x=$ 0.39 and 0.37. Solid lines are KTB lines for independent bilayers, dashed lines are KTB lines for coupled bilayers}
\label{fig:all}
\end{figure}

We start with overall look at the superfluid density of severely underdoped YBCO films, Fig.~\ref{fig:all}
\begin{figure*}[p]
\includegraphics[height=\textheight]{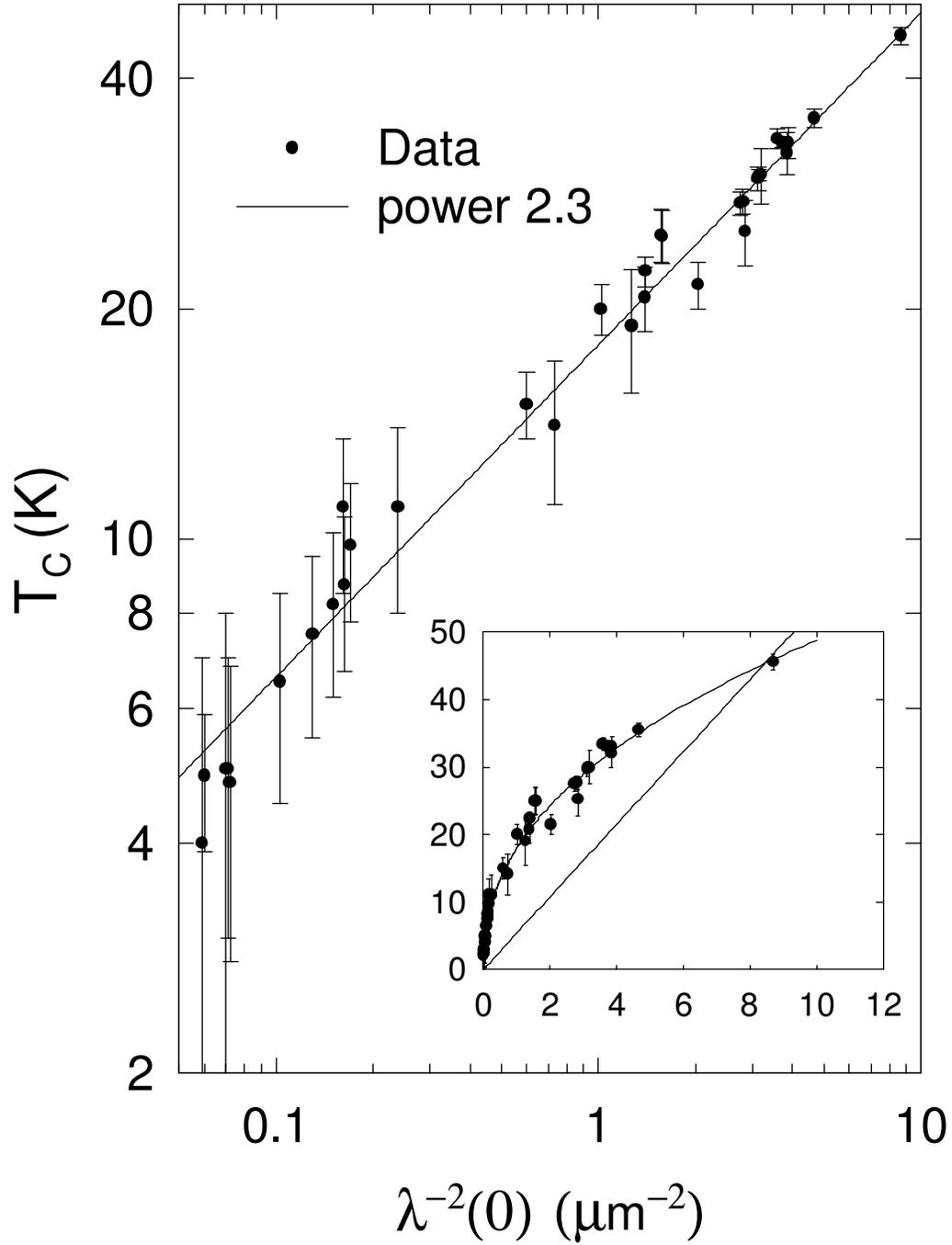}
\caption{Plot of $T=0$ superfluid density vs. $T_C$ in log-log (main graph) and linear-linear (inset) plots. The best power-law fit is $\lambda^{-2}(0)\propto T_C^{2.3}$. Vertical error bars are widths of $\sigma_1$ peaks. The solid straight line in the inset shows the $T_C$ from quasi-2D thermal phase fluctuations, Eq.~\ref{eq:KT}}.
\label{fig:Tcvsns}
\end{figure*}
This figure presents a series of films with $T_C$'s from 6 to 46 K. The narrowest transitions are about 2 K wide, as given by the width of the $\sigma_1$ peak. For purposes of determining $\lambda^{-2}(0) vs. T_C$, we consider these transitions sufficiently narrow because at the lowest temperatures of the experiment condition $\sigma_1 \ll \sigma_2$ is satisfied. Moreover, we have measured many more films than are presented here, and all gave the same results for $\lambda^{-2}(0) vs. T_C$, unless the width of the peak in $\sigma_1$ extended to $T = 0$. Films that were optimally oxygenated typically had $T_C$'s near 90 K, and $\lambda^{-2}(0) \approx 25 \mu m^{-2}$, and $\lambda^{-2}(T)$ was quadratic in $T$ at low T. These results are consistent with the films being slightly disordered d-wave superconductors. Assuming that the scattering rate does not change significantly with underdoping, and that the superconducting gap is the pseudogap when $T < T_C$, then the underdoped films are actually "cleaner" than optimally doped films in the sense that the scattering rate is a smaller fraction of the superconducting gap energy.
 
In quasi-2D layered superconductors we expect the observed transition to occur just a little above the 2D-XY transition temperature $T_{2D}$ of a single CuO$_2$ layer, which is related to the measured magnetic penetration depth by the familiar relation:

\begin{equation}
kT_{2D}=\frac{\Phi_0^2}{8\pi \mu_0}\frac{d}{\lambda^2(T_{2D})}
\label{eq:KT}
\end{equation}
where $\Phi_0$ is the flux quantum and $d=$11.7\AA is the thickness of 1 unit cell. (Note that if we use the film thickness in the above equation, we get an upper limit on the temperature at which thermal phase fluctuations must become important.) In fig.~\ref{fig:all}, intersections of the solid straight lines with the measured $\lambda^{-2}(T)$ curves give $T_{2D}$. Quite obviously, $\lambda^{-2}(T)$ does not vanish only a little above that temperature. In fact, for the two most severely underdoped films, the observed $T_C$'s are 5 times larger than $T_{2D}$. On the other hand, intersections of the dashed lines with $\lambda^{-2} (T)$ are the 2D transition temperatures predicted by using the full film thickness in the above equation. The so-derived $T_{2D}$'s closely match the positions of the peaks in $\sigma_1$. We conclude that the characteristic length for the Kosterlitz-Thouless transition is the film thickness and not a single unit cell thickness; severely underdoped films are not quasi-2D insofar as thermal phase fluctuations are concerned. Present results on many films augment our earlier report on the $T_C = 34$ Kfilm~\cite{Zuev}

We now turn to the central result of this report, which is the comparison of our data with the famous Uemura plot~\cite{Uemura}. Figure~\ref{fig:Tcvsns} shows $T_C$ {\em vs.} extrapolated values $\lambda^{-2}(0)$ in log-log and linear-linear (inset) scale. We have measured many more samples than appear in Fig. 1, and all of them fall on single line, irrespective of annealing procedure or transition width. In particular, samples with same oxygen content, but with different degree of Cu-O chain disorder (and hence different $T_C$ and $\lambda^{-2}(0)$ ) fall on the same curve as samples with different oxygen contents. It is clear that $T_C$ is not proportional to $n_s(0)$. Instead, we find $\lambda^{-2}(0)\propto T_C^{2.3 \pm 0.4}$. The straight line in the inset to Fig.~\ref{fig:Tcvsns} shows prediction for $T_C$ from quasi-2D thermal phase fluctuations, Eq. ~\ref{eq:KT}, with $d=$1.17nm. Below about 40 K the observed $T_C$ is larger than this upper limit. We conclude that thermal phase fluctuations are not responsible for the decrease in $T_C$ of underdoped cuprates.

Values of $\lambda^{-2}(0)$ in underdoped YBCO films are about 5 times smaller than is seen in the cleanest YBCO crystals with the same $T_C$ ~\cite{UBC2}. Nevertheless, both sets of samples reveal the same dependence of $T_C$ on $n_s(0)$. This is important because YBCO films are much more similar to other cuprates, e.g., BiSrCaCuO and LaSrCuO, than are YBCO crystals. Ultraclean YBCO crystals are special. At optimal doping, they possess the highest superfluid density of all hole-doped cuprates by a factor of 4.

Our central result is the finding that $\lambda^{-2}(0)\propto T_C^{2.3 \pm 0.4}$ for heavily underdoped YBCO films. This result disagrees with expectations based on the idea of cuprates as quasi-2D layered superconductors and with the phenomenological proportionality, $\lambda^{-2}(0)\propto T_C$, implied by the Uemura plot. Regarding the latter, it is worth noting that most of the data in the original Uemura plot come from samples that are not as severely underdoped as the samples presented here. A secondary result is that $T_C$ is not limited by quasi-2D thermal phase fluctuations, as they are understood from simulations of Josephson-coupled superconducting grains. Coupling between CuO$_2$ planes is apparently strong enough to make fluctuations firmly 3D and therefore relatively unimportant up to temperatures few K below $T_C$. 

Why does $\lambda^{-2}(0)$ decrease so rapidly with underdoping? If the pseudogap is due to an order parameter unrelated to superconductivity, like the $d$-density wave order parameter of Chakravarty et al.,~\cite{chakravarty} then one can appeal to disorder (scattering) and a decreasing gap, $\Delta \propto T_C$ to account for the reduction in $\lambda^{-2}(0)$ with underdoping. If the pseudogap is the superconducting gap, then the films become cleaner with underdoping, and scattering cannot be the explanation. It is possible that percolation of some sort is important. If superconductivity is confined to localized regions, like malformed stripes, then the measured superfluid density may be determined by coupling between regions. Finally, it has been speculated that electronic charge is renormalized to zero away from $d$-wave nodes in underdoped cuprates~\cite{UBC1,UBC3}. In the end, currently there is no reliable model that we are aware of, that predicts or explains our finding.

 In conclusion, at severe underdoping critical temperature and superfluid density are not proportional. Instead, they are related by power law: $n_s(0)\propto T_C^{2.3}$. This is contrary to understanding of the pseudogap as a suppression of $T_C$ due to thermal phase fluctuations.


\begin{thebibliography}{10}

\bibitem{PLee}P. A. Lee, cond-mat/0307508

\bibitem{Uemura} Y. J. Uemura {\em et al.} Phys. Rev. Lett. {\bf 62}, 2317 (1989)

\bibitem{KTB} J.M. Kosterlitz and D.J. Thouless, J.
Phys. C {\bf 6}, 1181 (1973); J.M. Kosterlitz, {\it ibid} {\bf 7},
1046 (1974); V.L. Berezinskii, Sov. Phys. JETP {\bf 32}, 493 (1971).

\bibitem{EmeryKivelsonPRL}V. J. Emery and S. A. Kivelson, Phys. Rev. Lett, {\bf 74}, 3253, (1995)
\bibitem{EmeryKivelsonNature}V. J. Emery, S. A. Kivelson, Nature, {\bf 374}, 434-437 (1995)

\bibitem{Carlson1}E. W. Carlson, S. A. Kivelson, V. J. Emery, , E. Manousakis, Phys.Rev. Lett, {\bf 83}, 612, (1999)

\bibitem{CurtyBeck}P. Curty and H. Beck, Phys Rev. Lett. 91, 257002 (2003).

\bibitem{Muller}D. Mihailovi{\'c}, V.V. Kabanov, and K.A. M{\"u}ller, Europhys. Lett. 57, 254-259 (2002).
\bibitem{IoffeMillis} L.B. Ioffe and A.J. Millis,  J. Phys. Chem. Solids 63, 2259-2268 (2002).

\bibitem{Scalapino}T. Eckl, D.J. Scalapino, E. Arrigoni, and W. Hanke, Phys. Rev. B 66, 140510(R) (2002).

\bibitem{Timm} C. Timm, D. Manske, and K.H. Bennemann, Phys. Rev. B 66, 094515 (2002).

\bibitem{Hirschfeld} H.J. Kwon, A.T. Dorsey, and P.J. Hirschfeld, Phys. Rev. Lett. 86, 3875-3878 (2001).

\bibitem{Tesanovic}Z. Tesanovi{\'c},  Phys. Rev. B 36, R2364 (1987).

\bibitem{Jorgensen} J. D. Jorgensen, B. W. Veal, A. P. Paulikas, L. J. Nowicki, G. W. Crabtree, H. Claus and W. K. Kwok, Phys. Rev. B {\bf 41},1863 (1990)

\bibitem{chakravarty} S. Chakravarty, R.B. Laughlin, D.K. Morr, and C. Nayak,
Phys. Rev. B {\bf 63}, 094503 (2001).
 
\bibitem{2coil1} S. J. Turneaure E. R. Ulm and T. R. Lemberger, J. Appl. Phys. {\bf 79}, 4221, 1996

\bibitem{2coil2} S. J. Turneaure, A. A. Pesetski and T.R. Lemberger,  J. Appl. Phys.,{\bf 83}, 4334, 1998

\bibitem{Zuev}Y. Zuev, J. A. Skinta, M.S. Kim, T. R. Lemberger, E. Wertz, K. Wu, and Q. Li, cond-mat/0407113

\bibitem{UBC2}D. Bonn, Private Communication.

\bibitem{UBC1} A. Hosseini, D. M. Broun, D. E. Sheehy, T. P. Davis, M. Franz, W. N. Hardy, R. Liang, and D. A. Bonn, Phys. Rev. Lett. {\bf 93}, 107003 (2004)

\bibitem{UBC3}D. E. Sheehy, T. P. Davis, and M. Franz, Phys. Rev. B {\bf 70}, 054510 (2004)



\end{thebibliography}
\end{document}